\documentclass{aa}
\usepackage{graphicx}
\usepackage{txfonts}
\usepackage{natbib}
\usepackage{aalongtable}
\newcommand{\kms}{km\,s$^{-1}$}
\newcommand{\mkms}{\rm km\,s^{-1}}
\bibpunct{(}{)}{;}{a}{}{,}

\begin{document}
		\title{Accurate stellar rotational velocities using the Fourier 
		transform of the cross correlation maximum}
   	\author{C. G. D\'{\i}az \inst{1}, J. F. Gonz\'alez \inst{1,2}, 
   	H. Levato \inst{1,2}, \and M. Grosso \inst{1,2}}
		\institute{Universidad Nacional de San Juan - Av. J. I. de la 
		Roza 590 oeste, 5400 Rivadavia, San Juan, Argentina 
		\\ \email{cgonzadiaz@gmail.com}
\and ICATE - CONICET - Av. Espa\~na 1512 sur, J5402DSP San Juan, Argentina  
\\ \email{fgonzalez@icate-conicet.gob.ar, hlevato@icate-conicet.gob.ar}
}
		\date{Received ; accepted }
	\abstract
		{}
	{ We propose a method for measuring the projected rotational velocity 
	$v sin i$ with high precision even in spectra with blended lines. 
	Though not automatic, our method is designed to be applied systematically
	to large numbers of objects without	excessive computational requirement.
}
	{We calculated the cross correlation function (CCF) of the object spectrum 
	against a zero-rotation template and used the Fourier transform (FT) of the 
	CCF central maximum to measure the parameter $v\sin i$ taking the limb 
	darkening effect and its wavelength dependence into account . The procedure 
	also improves the definition of the CCF base line, resulting in errors 
	related to the continuum position under 1 \% even for $v\sin i$ = 280 \kms.
	Tests with high-resolution spectra of F-type stars indicate that an 
	accuracy well below 1\% can be attained even for spectra where most lines 
	are blended.
}
	{ We have applied the method to measuring $v\sin i$ in 251 A-type stars. 
	For stars 	with $v\sin i$ over 30 \kms\,(2--3 times our spectra resolution), 
	our measurement errors are below 2.5\% with a typical value of 1\%.  We compare 
	our results with Royer et al. (2002a) using 155 stars in common, finding 
	systematic differences of about 5\% for rapidly rotating stars. 
}
		{}
\authorrunning{C. G. Diaz et al.}
\titlerunning{Accurate stellar rotational velocities based on CCFs}
 \maketitle
\section{Introduction}
The projected axial rotational velocity of single stars can be measured directly from 
the broadening of their spectral lines. That is why the rotational velocity is an 
important observable in statistical studies of stellar astrophysics. Several methods 
have been developed to measure $v\sin i$, but the problem of systematic differences 
between different authors and methods has always been present. The first workable model
for the stellar rotation was established by \citet {carroll28,carroll33} and 
\citet {caringra33}, but \citet {stru29} presented a very simple graphical model called the
classical model of a rotating star (CMRS) by \citet{ct95}, which was used as a standard 
for most of the work done in the 20th century  \citep[for more details see the excellent 
review paper by][]{col04}. \citet {stru29} did not consider limb darkening, which was 
introduced in the CMRS by \citet {carroll33}.
In a series of papers Slettebak measured $v\sin i$ for stars in the spectral range O-G 
\citep{sle54,sle55,sle56,sleho55} by considering a linear limb darkening law. 
These stars were used in many subsequent papers by different authors to calibrate the full 
width at half depth of the line used to measure the $v\sin i$ parameter. 
Therefore, determination of $v\sin i$ was conditioned by the calibration made by the author. 

In the last quarter of the 20th century \citet{sl75} established a new standard system of 
stars distributed over both hemispheres and covering a range in spectral type from O9 to F9. 
The measure of $v\sin i$ was again based on a calibration of $v\sin i$ with the full width 
at half depth of stellar absorption lines in the star spectrum. 
\citet {sl75} established this system by comparing their digital data with numerical models 
constructed by adopting Roche geometry, uniform angular velocity at the surface, von Zeipel
gravity darkening, and numerical integration of angle-dependent model atmosphere intensities. 
The standards for the high-velocity specimens were established by \citet{sle82} by visual 
comparison of line widths on the spectrograms that were recorded on photographic plates. 
Systematic effects between the old and new Slettebak systems were studied by \citet{gar84}. 
The calibrations of the systems described were made for only three lines, \ion{He}{i} 4471 
{\AA}, \ion{Mg}{ii} 4481 {\AA}, and \ion{Fe}{i} 4476 {\AA}, and the theoretical profiles were 
calculated with main sequence models. 
Besides this, determination of the full width at half depth has shown to be very sensitive to 
the continuum position, which represents an important error source.

A significant improvement was possible thanks to methods based on Fourier transform (FT) 
of line profiles providing a deeper analysis. In the first place, this tool is used to 
avoid an external calibration by suppressing the error related to this stage of the process. 
In second place, with high signal-to-noise ratio (S/N) it is possible to identify another broadening 
agents present in the line profile, at least in a qualitative way. 
Finally, it is possible to analyze second-order effects like differential rotation 
\citep{g77,g82,b81,rs02} and to put a limit on the inclination of the rotational axes 
\citep {rr04}. Regardless of the method adopted to determine $v\sin i$, it is not an easy task to 
find the spectral lines that have the minimum conditions required to be used in the measurement 
of rotational velocity, namely: 1) lines with no blends, 2) lines intense enough to be identified in 
rapid rotators, and 3) lines mainly broadened by rotation. 
For instance, in O-type and B-type stars, only the Balmer lines and some helium lines are 
intense enough, but they all show a significant Stark effect. 
In stars of spectral type later than A4, almost all lines are blended  if $v\sin i$ is grater 
than 100 \kms, making it impossible to find an isolated line.
This problem is evident in the work of \citet{roy02a} where fewer than three lines were measured 
in A-type stars with $v\sin i > 60$ \kms. 
Because this problem increases with the spectral type, other methods have been developed. 

One solution to the blending limitation is the use of a least square deconvolution procedure 
to derive the broadening function in a selected wavelength region instead of a single line 
profile. This methodology implies the application of an iterative process to fit the equivalent 
width of the template's spectral lines, the broadening function, and the continuum position 
\citep [for details see][]{rs03}. 
This is a very powerful technique for a detailed study of the rotational profile.
However, for extensive applications like the development of a catalog
of rotational velocities, a more direct method that does not involve
fitting the intrinsic spectrum or including any atmospheric parameter other 
than limb darkening, might be more suitable.

Even though CCF had been originally proposed for determining radial velocities, the projected 
axial rotational velocity can be inferred from the width of the CCF maximum. 
This tool has often been applied to determine $v\sin i$ in cool stars (M-L spectral type). 
The standard procedure is to calculate the CCF between an observed spectrum and a template 
spectrum of the same temperature with $v\sin i$ $=$ 0 \kms. 
Then, a fitting profile for the maximum of this function is calculated and the width of the
fit can be used to measure $v\sin i$ through an empirical calibration $v\sin i$-width.
The absence of single lines is somehow solved by means of the CCF .
Nevertheless, as in any other method that depends on empirical calibrations, various 
strategies, not always equivalent, have been adopted in the literature.
In some works the central maximum is fitted with a Gaussian \citep{bj04}, while in others 
a parabola \citep{tr98},  or even a Gaussian plus a quadratic function are used \citep{wb03}.
The adoption of different functions to evaluate the rotational broadening 
could lead to systematic differences in results from different authors.

Templates selection is also very heterogeneous. Even though some authors calculate synthetic 
template spectra, real stellar spectra have been used in most works based on the CCF  
\citep[see][]{wb03,mb03,bj04}.
Since the CCF contains information from the template and the object spectrum, using an observed 
template could introduce an external error source from the unknown broadening factors present 
in the template spectral lines. Nowadays, the best methods dealing with stars that show intense 
line blending in their spectra require significant computational resources. 

Motivated by the need for a precise and expeditious technique to be applied extensively 
for the construction of a catalog of rotational velocities of bright A-type stars (Levato 
et al. in preparation), we develop here an alternative method based on the CCF and, at the 
same time, independent of any external calibration. 
Our procedure uses the FT to measure the parameter $v\sin i$, taking 
the dependence of the transform with limb darkening into account.
In Sect.\,2 a full description of the methodology is presented. 
Limb darkening consideration and other practical details of the procedure are explained 
in Sect.\,3. Sect.\,4 describes specific application to A and later-type stars. 
The precision of the obtained rotational velocities is discussed  in Sect.\,5,  
and our main conclusions are summarized in Sect.\, 6.

\section{Method}

As a first approximation, according to the classical model of a rotating star revised by 
\citet{ct95} \citep[see also][]{bv97}, we can consider an observed stellar line profile 
as the convolution of an intrinsic line profile and a rotational broadening function. 
Then, the observed spectrum $D(\lambda)$ can be approximated by a convolution between a 
template  spectrum $T(\lambda)$ and the rotational broadening function $G(\lambda)$,
$$D(\lambda)= T(\lambda)\ast G(\lambda),$$ where $T(\lambda)$ includes any other broadening 
effect different from rotation: natural line broadening, thermal broadening, microturbulence, 
Stark effect, etc. 

Under the previous assumption the CCF between the object spectrum and a template of the same 
spectral type, but without rotation, results in
\begin{equation}\label{eq1}
CCF_{DT} = \frac{D \ast T}{\parallel D\parallel \cdot \parallel T\parallel} = \frac{(T \ast G) \ast T}
{\parallel T\parallel \cdot \parallel T\parallel} = CCF_{TT} \ast G,
\end{equation}
where $\parallel T\parallel$ and $\parallel D\parallel$ are the norm of each spectrum.
Since the rotational velocity is the main broadening agent of metallic lines in stars 
with $v\sin i > 10$\,\kms, i.e. in most normal main-sequence stars \citep{am95,alg02}, 
the intrinsic width of a line profile $\sigma$ is usually $\sigma \ll v\sin i$.

The autocorrelation function $CCF_{TT}$ presents a narrow peak centered at zero, 
which for the current application can be considered as a Gaussian whose width is 
$\sqrt 2$ times the width of the template lines.
Side lobes are present in this function, but they are usually very small in comparison
with the central peak since their intensity is inversely proportional to the number 
of spectral lines involved  in the cross-correlation.
Therefore, regardless the shape of the line profile of the template, the function 
$CCF_{DT}$ is very similar to the rotation function $G$ and can be used to derive the 
parameter $v\sin i$.

The idea of the proposed method is to calculate the FT of the central maximum of the
$CCF_{DT}$ to derive $v\sin i$ from the position of its first zero.
In fact, $CCF_{DT}$ is essentially the function $G$ convolved with a narrow profile that
has no significant impact on the position of the first zero of the FT.
The central maximum of the CCF has the same information about the rotational velocity 
as a single line profile. The main benefit of utilizing the CCF is that having to select 
lines without blending, which represents a problem in stars later than late-A 
early-F, is avoided.

The CCF peak, however, might be blended with side lobes, but the more
lines used to calculate the CCF, the bigger the intensity difference  between the central 
maximum and the secondary maxima.
In the proposed method, we include a procedure to remove the side lobes contribution as 
described in Sect.\,\ref{sect.extrotpr}.

Finally, the S/N of the CCF peak is much higher than that of a single line.
Therefore, using the CCF calculated from a large spectral region instead of individual 
line profiles presents advantages over both line blending and S/N.
Nevertheless, there are some restrictions on the size of the spectral region to be 
correlated owing to the variation in the limb darkening coefficient with wavelength, which
will be discussed in the next section.

\subsection{Extraction of the rotational profile}\label{sect.extrotpr}

The $CCF_{DT}$ would be equal to the rotational profile $G$ only in the ideal case 
in which $CCF_{TT}$ has a single peak of negligible width (a Dirac $\delta$ function).
In practice, however, besides the finite width of the central peak, the $CCF_{DT}$ contains 
several small subsidiary peaks coming from the fortuitous coincidence of each spectral 
line of the object spectrum with different spectral lines in the template as the former is 
shifted with respect to the latter.
These small peaks can be blended with the central peak due to the rotational broadening of 
the object, affecting the determination of the base line of the CCF central maximum and hence 
the measurement of the rotational velocity.

\begin{figure}[t]
	\centering
\includegraphics[width=1\linewidth]{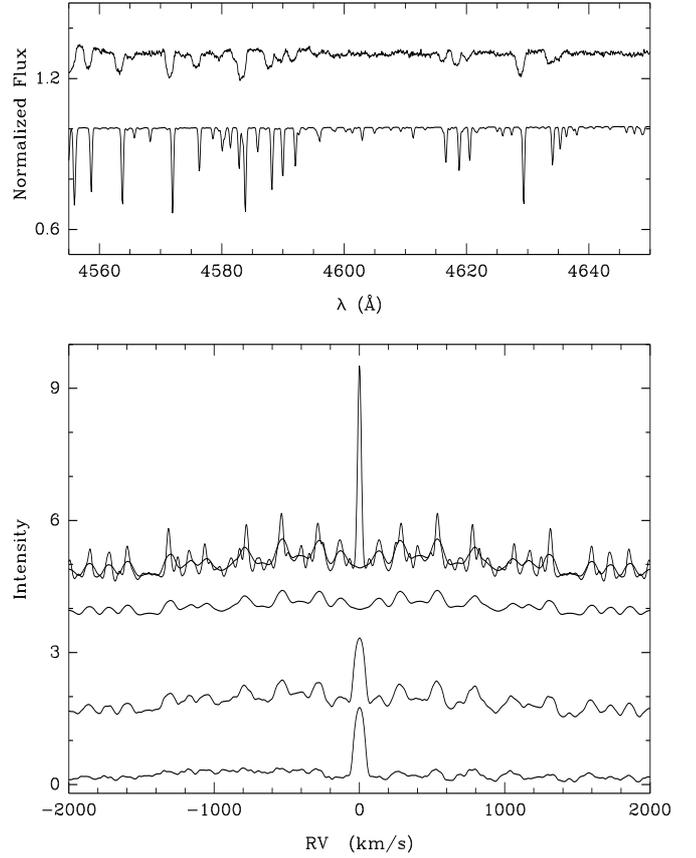} 
\caption{Calculation of the rotational profile using 
cross-correlations. Upper panel: Object spectrum and template spectrum. 
Lower panel, from top to bottom: a) Cross-correlation function template-template 
$CCF_{TT}$ (thin line) and the same function after removing the main peak and 
convolving with the provisional rotational profile, i.e. $CCF2_{TT}\ast G$ 
(thicker line), b) The same function $CCF2_{TT}\ast G$ scaled to the object 
intensity, c) Cross-correlation function object-template $CCF_{DT}$, and d) 
Rotational profile calculated subtracting $CCF2_{TT}\ast G$ from $CCF_{DT}$. 
In both panels arbitrary vertical shifts have been applied for clarity.}
\label{fig:ccf1}
	\end{figure}

To overcome this problem, a key stage in the process is the remotion of secondary 
maxima. To this aim we consider the function $CCF_{TT}$ as the sum of two functions: 
$CCF1_{TT}$, representing only the central peak, and $CCF2_{TT}$, which includes all the 
subsidiary peaks. Then, from eq. \ref{eq1} 
$$CCF_{DT} = CCF_{TT} \ast G = CCF1_{TT} \ast G + CCF2_{TT} \ast G.$$
To calculate the contribution of the subsidiary peaks, we remove the central maximum 
from the $CCF_{TT}$ to obtain $CCF2_{TT}$, which is convolved with a first approximation of the 
rotational broadening profile $G_{1}$ extracted from the central peak of  the $CCF_{DT}$. 
The result is then subtracted from the $CCF_{DT}$, thereby improving the determination of the base of 
the rotational profile. 
Considering the central peak of the $CCF1_{TT}$ to be much narrower than G, we find from the 
previous equation
\begin{equation}\label{eq.Gcor}
G \approx   CCF_{DT} - CCF2_{TT} \ast G_{1}.
\end{equation}
Before applying the subtraction, two corrections are required: a radial velocity correction for  
both CCFs to be centered,  and a scale correction to account for an eventual global line 
intensity difference between the object and the template spectra.
As illustration, Figure\,\ref{fig:ccf1} shows the application of this process to a spectrum
of the A2-type star HR\,892 with a synthetic template for $T=9000$\,K.

Finally, once the base line of the central maximum has been improved, we fit the 
background in the surrounding region of this peak and subtract the fit to have the 
base of the rotational profile at zero. Through this process the rotational 
broadening profile $G$ is obtained.

\begin{figure}[t]
	\centering
\includegraphics[width=1\linewidth]{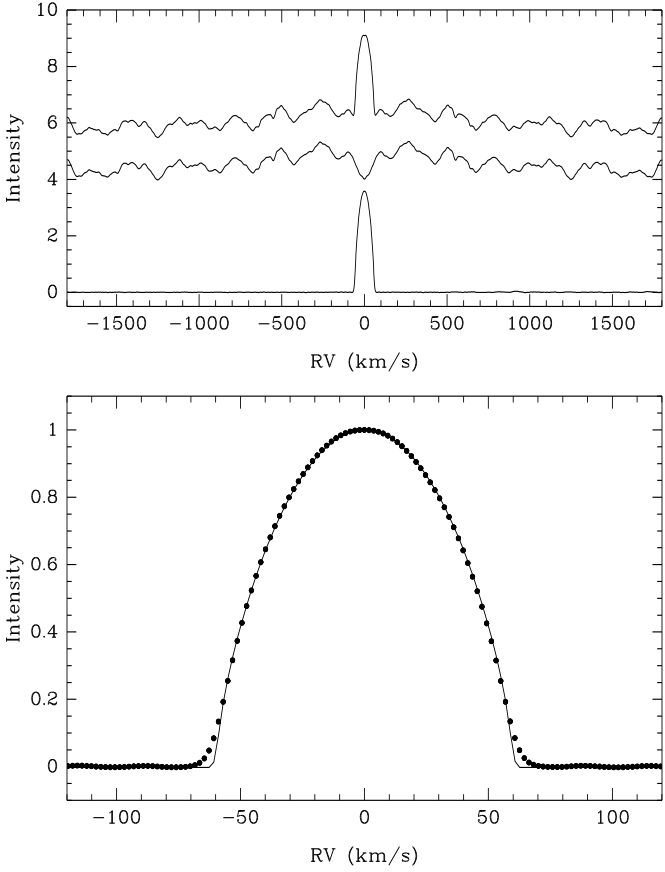}
\caption{Rotational profile recovered from the CCF. Upper panel:
the functions involved in equation \ref{eq.Gcor}, from top to bottom: 
CCF$_{DF}$, CCF2$_{TT}$*G$_1$, and the  recovered rotational profile $G$.  
Lower panel: comparison of the retrieved rotational profile 
(filled circles) with the input theoretical profile (solid line).}
\label{plot5400}
	\end{figure}

In principle, reliable and precise rotational profile can be reconstructed
from the CCF, as long as the rotational broadening is not comparable to
other photospheric or instrumental broadening effects.
In Fig. \ref{plot5400} we compare the theoretical rotational profile of eq. 
\ref{eq.G} with the profile recovered from the CCF of a 100\,\AA\, long 
spectrum (5350--5450 \AA).
As shown in the lower panel, the only noticeable difference is the smoothing 
of the sharp cut at the edges, owing to the intrinsic spectral line width 
(0.08--0.16 \AA).
The example corresponds to an atmosphere model with T$_{eff} = 8000$\,K and 
log\,$g$=4.0, convolved with a rotational profile of $v\sin i = 60$\,\kms and 
$\varepsilon = 0.6$. 

\subsection{Calculation of  $v \sin i$}

The rotational profile of a spectral line centered on a wavelength $\lambda_o$ for
a spherical star rotating as a rigid body and whose limb darkening law is linear 
with a coefficient $\varepsilon$, is
\begin{equation}\label{eq.G}
G(x) = \frac{2(1-\varepsilon)(1-x^2)^{1/2}+\frac{\pi\varepsilon}{2} (1-x^2)}{\pi (1-\frac{\varepsilon}{3})}
\end{equation}
for $x<1$, and $G(x)=0$ for $x>1$, where $x=\ln(\lambda/\lambda_o)\cdot c/(v\sin i)$.
Expressing the rotational profile in terms of $\Delta \ln \lambda$ instead of
$\Delta \lambda$ has the advantage of being independent of the line wavelength 
 (cf. with eq.\,17.12 in \citealt{g92}; eq.\,4 in \citealt{rs02}).
In addition, the Doppler formula in logarithmic scale, $\Delta \ln\lambda 
\approx \mathrm{v}/c$, is a better approximation than the classical $\Delta \lambda/\lambda 
\approx \mathrm{v}/c$.

The FT of $G(x)$ presents zeros at different positions $\sigma_{n}$, which are related 
to $v\sin i$ by
\begin{equation}
v \sin i = \frac{k_n(\varepsilon)}{\sigma_n}, \label{eq:vs1} 
\end{equation}
where $k_n(\varepsilon)$ are functions of the limb darkening coefficient \citep{rs02}.
For the first zero, $k_1$ is given implicitly by the following expression:
$$
\frac{1}{\varepsilon} = 1 + \left[ \frac{\sin\beta}{\beta^2}-\frac{\cos\beta}{\beta}\right]\frac{1}{J_1(\beta)}
$$
where $J_1$ is the first-order Bessel function and $\beta = 2\pi k_1$.
In our procedure we used the following approximate formula
$$
k_1 = 0.60975 + 0.0639 \varepsilon + 0.0205 \varepsilon^2 + 0.021  \varepsilon^3,
$$
which is accurate enough (better than 0.01\%) for the whole range of limb darkening
coefficient of normal main-sequence stars: $\varepsilon$ = 0.00--1.10.
We note that this cubic is not a series expansion around $\varepsilon=0$ but a fit 
of the function in this interval. In fact, it is more accurate than the fourth-degree
polynomial calculated by 
\citet[][ eq. 6; eq. 8 in \citealt{rs02}]{drav90} 
in the range of stellar limb darkening coefficients.

Once $k_1(\varepsilon)$ has been calculated, the determination of $v\sin i$ results 
from eq.\,\ref{eq:vs1}. We calculate the power spectrum of the extracted rotational 
broadening function to evaluate the position of the first zero  $\sigma_1$ of the FT 
and calculate $v\sin i$ from eq.\,\ref{eq:vs1}.
As mentioned before, this procedure is valid for stars rotating rigidly. The eventual 
presence of differential rotation might be detected by studying the shape of the derived  
rotational profile or using the first two zeros of the FT \citep{rs02}.

Finally, it is important to mention that a linear limb darkening law was used in this work, but it
is possible to consider other types of limb darkening law like the ones compared by \citet{bv97}.
We estimate that a linear law could produce errors of $\sim 1\%$; therefore, a more adequate 
limb darkening 
law is a potential improvement for the future.

\section{Procedure and assumptions}
The procedure was programmed as an IRAF task divided in five stages:
\begin{enumerate}
\item calculation of the CCF between an object spectrum and a template spectrum,
\item extraction and cleaning of the CCF central maximum,
\item determination of the first zero of the FT of the CCF maximum,
\item calculation of  $v\sin i$,
\item measurement error estimation.
\end{enumerate}
In this section we describe four important aspects for implementing our method.

\subsection{Influence of nonrotational broadening effects}

\begin{figure}[t]
	\centering
\includegraphics[width=1\linewidth]{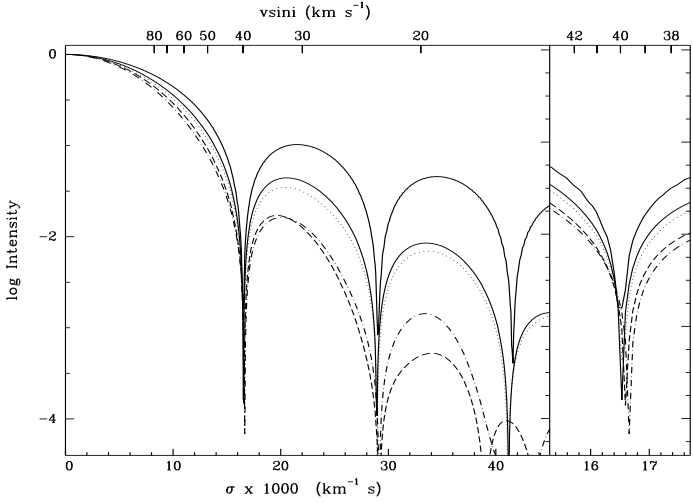} \caption{ 
Fourier transform of the CCF of 5 synthetic spectra ($\Delta \lambda = 120$\AA) with 
$T_\mathrm{eff}=9000$ K,  $v\sin i=40$\, \kms\, and convolved with the following 
Voigt profiles: 
Thick solid line: original rotational profile.
Thin solid line: $\rm{gfwhm}=0.2\AA$.
Dotted line: $\rm{gfwhm}=0.1\AA$ and $\rm{lfwhm}=0.1\AA$.
Dashed line: $\rm{gfwhm}=0.4\AA$. 
Dot-dashed line: $\rm{gfwhm}=0.2\AA$ and $\rm{lfwhm}=0.2\AA$.}
\label{powlog40}
	\end{figure}

We evaluated the impact of nonrotational broadening effects
on the position of the first zero and on the shape of the main lobe of the FT
using a synthetic spectrum for a $T_\mathrm{eff}=9000$ K atmosphere model in the 
wavelength range 4480 -- 4600 \AA. 
This spectrum was convolved with a rotational profile of $v\sin i=40$\, \kms\, 
and different Voigt profiles to simulate additional nonrotational broadening.
The FT of the various broadening profiles, derived using the CCF, are
compared in Figure \ref{powlog40}. 
We used Voigt profiles corresponding to different combinations of Gaussian 
(gfwhm) and Lorentz (lfwhm) components' full-width-half-maximum (FWHM).
The change in the general shape of the first lobe with the nonrotational broadening 
effects is more significant than the shift of the first root, which 
in the four examples lies within $0.4$ \kms\, from the original rotational profile.
This has been already noted by \citet{rs02}. 

We estimate that acceptable $v\sin i$ measurements are obtained as long as the rotational profile
is at least twice wider than the nonrotational profile.
In main-sequence stars, typically the intrinsic FWHM of metallic lines
is 0.07--0.12 \AA. Therefore, for high-resolution spectra, the lower limit
for $v \sin i$ measurements through the zero of the FT is about $v\sin i=$5--8 \kms.
B-type stars are more difficult to measure since their most conspicuous 
spectral features are \ion{He}{i} lines, whose intrinsic widths are usually 
on the order of 0.8 \AA. Such lines would not be suitable for rotation measurement 
with this technique, except for very fast rotators ($v\sin i \gtrsim 100$\,\kms).
In fact, these lower limits depend on the S/N, since the effect of the intrinsic 
line broadening is to reduce the intensity of the first subsidiary lobe, making it
more difficult to determine the first zero position.

\begin{figure}[t]
	\centering
\includegraphics[width=1.\linewidth]{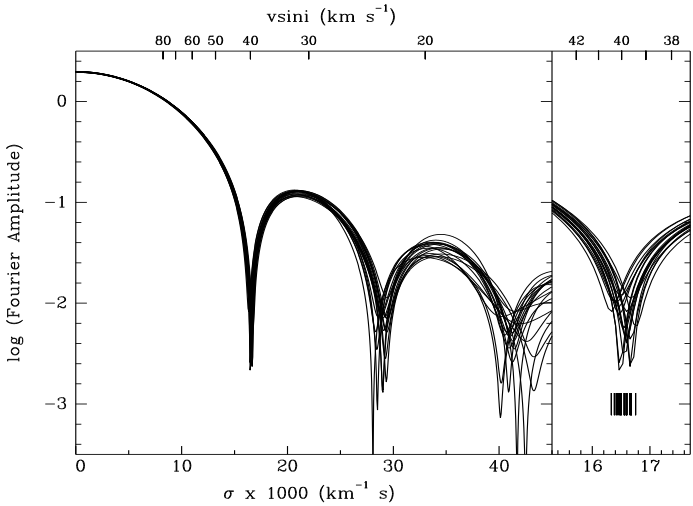} \caption{
Fourier transform of the CCF of 20 synthetic spectra ($\Delta \lambda = 120$\AA) with 
$T_\mathrm{eff}=9000$ K,  $v\sin i=40$\, \kms\, and random noise
simulating S/N = 100.
 }
\label{powlog100}
	\end{figure}

All in all, for the usual values of S/N and instrumental broadening,
the variation in the first zero position caused by additional broadening
and noise is below 1\%. A typical case is shown in 
Fig.\ref{powlog100}, where the FT of 20 spectra with different 
random noise corresponding to $S/N = 100$ are plotted. 
The atmospheric parameters, $v\sin i$, and spectral range are the same as 
for Fig. \ref{powlog40}, and the assumed instrumental profile is 6.6 \kms.
We measured $\langle v\sin i\rangle=39.88$\,\kms with $\sigma =0.26$\,\kms ($0.7\%$) 
for $S/N=100$. Using $S/N=50$ the typical error is $\sigma = 1.3\%$.
Considering that we used just a 120 \AA\,wide region, we conclude that,
with medium quality spectra (e.g. $\Delta \lambda \approx 1000$ \AA\, and 
$S/N=50$ or $\Delta \lambda \approx 100-200$ \AA\,and $S/N=100$), errors are 
well below 1\%.

In light of these results and under the assumption of rigid rotation,
we consider that a global fitting of the FT of the rotational profile does not 
present much of an advantage for measuring rotational velocity, since the 
general shape of the function depends on the nonrotational broadening effects
much more strongly than the position of the first root does, and therefore, 
additional free parameters should be included in the fitting in order to 
model the nonrotational contributions simultaneously.
Moreover, even though the S/N is evidently lower around the first root, 
it is usually high enough for this purpose, since the CCF peak has a much 
higher S/N than individual spectral lines.

We note that the FT zero position is even less sensitive to nonrotational 
broadening than the Bessel-Fourier transform proposed by \citet{pgp96}. 
In fact, according to Fig.\,2 of Piters et al. (1996), the maximum of
the Bessel-Fourier transform is shifted by about 1\% (2\%) when an additional 
Gaussian broadening of gfwhm $= 0.14 \times v\sin i$ ($0.28\times v\sin i$) 
is present. These values approximately correspond to 0.2 and 0.4 \AA\ in the 
calculations of our Fig.\,\ref{powlog40}, for which we found a shift in the 
FT zero of only 0.2\% (0.4\%).

\subsection{Considerations on limb darkening}

The influence of the limb darkening coefficient on the rotational broadening 
function $G(\lambda)$ is well known  \citep[see][]{rs02,ct95}.
Although the errors can rise to nearly 16 \%, most authors assume a fixed value 
of $\varepsilon =$ 0.6 \citep {sh07,s06,rr04,roy02a,roy02b,ct95,r89}, even for
low-mass stars (spectral types M or L) and brown dwarfs \citep{wb03,mb03,bj04,tr98}.

It is also worth noting that the limb darkening coefficient in the core of the 
lines may differ substantially from that of the nearby continuum. \citet{ct95} 
treated this problem and indicate that the difference may be important. 
For a nonrotating B9V model they calculated $\varepsilon = 0.57$  for the 
continuum at $\lambda =4475$\AA \, and $\varepsilon = 0.29$ for the core of the 
\ion{Mg}{ii} line at $\lambda4481.13$. 
Even though the limb darkening coefficient varies within the intrinsic line profile,
it is valid to define the rotational profile $G(x)$ as in eq.\,\ref{eq.G} using
an effective limb darkening coefficient (e.g. weighted average over the line profile),
provided the rotational broadening is much larger than the intrinsic line profile. 
Our method does not use single lines but spectral regions that include absorption 
lines with $\varepsilon$ values that differ from the values of $\varepsilon$ for
the continuum in different amounts. 
This can still be considered by defining an effective limb darkening coefficient for 
each spectral range. However, we do not consider this issue in the present paper.
Instead, we split the spectrum to perform the measurements in regions of about 
200--400 \AA \, and adopt the limb darkening coefficient for the continuum at the 
central wavelength of each region as representative of the spectral region.

Moreover, owing to the dependence of $\varepsilon$ on temperature and wavelength,
the parameter $k_n(\varepsilon)$ in eq.\,\ref{eq:vs1}  varies with spectral type
and along the spectrum.
To account for this effect, we include a calibration $k_1(\lambda,T_{\rm eff})$ 
(see next section) specifically obtained and tested for A type stars. 
The purpose of having an empirical calibration for a particular range of temperatures 
or spectral type is to maximize the precision and to keep the procedure as simple 
as possible. Another effect that has not been considered is the gravity 
darkening, which is important for $v \sin i$ larger than 100 \kms. \citet {ct95} 
also discusses this effect and warns that the error may be as large as 10\% 
for $v \sin i$ larger than 200 \kms. 
Consideration of this effect may also be a second step towards improving the method.

\subsection{Template spectrum and spectral region selection}

\begin{figure}[t]
	\centering
\includegraphics[width=1\linewidth]{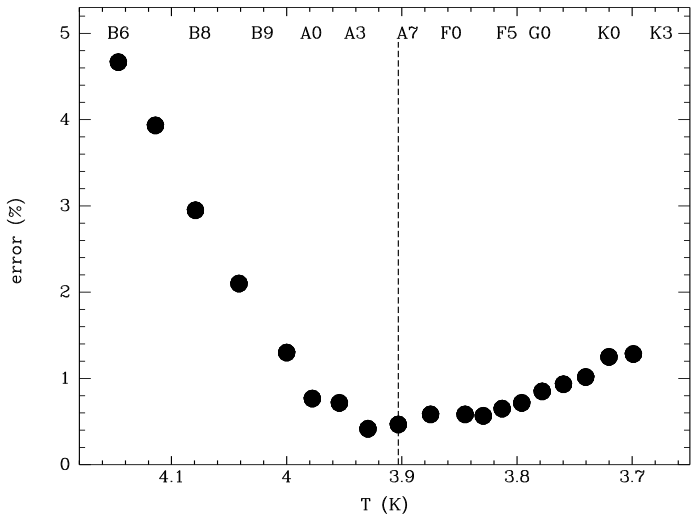} 
\caption{Error in $\langle v\sin i\rangle$ from the mismatch of spectral type 
between the object and the template spectrum. The $x$-axis has the temperature
of the template. 
The temperature of the object is indicated with the dashed line 
($T_\mathrm{eff}=8000$ K).}
\label{temp-err}
	\end{figure}

The template spectrum must be morphologically similar to the object spectrum 
but with zero rotational velocity. As the maximum of the CCF is the result of 
the contribution of all the coincident spectral lines between
both spectra, slight differences resulting from small discrepancies in spectral
type, metallicity, or the presence of spectral peculiarities have little influence 
on the CCF maximum.
We tested the incidence of using templates that differ morphologically 
from the target spectrum, measuring a synthetic spectrum for an A7V star 
($T_\mathrm{eff}=8000$ K and log\,$g$=4.0) and $v\sin i=60$\, \kms with various 
templates. We measured six spectral regions $\geq 240$\,\AA\, in the range 
3985 -- 5870 \AA\, with 19 main sequence templates ranging from 
$T_\mathrm{eff}=$5\,000 to 14\,000 K, and we found that the error produced by spectral
type mismatch is below 1\% for templates in the range A0 -- G0 
($T_\mathrm{eff}=$6 000 -- 10 000 K) as shown in Figure \ref{temp-err}. 
Although the present work is focused on A-type stars, from Fig. \ref{temp-err} it is 
also evident that earlier spectral types are more sensitive to this effect,
so this analysis should be carried out for the spectral range of interest in future 
works. 

We also used an A5V template to measure A0 and A9 stars in our program 
and a dwarf template to measure giants and supergiants of the same type. 
In all cases, the differences in the $v\sin i$ are always below 1\%. 
Such a small error shows the strength of the CCF, and that spectral type mismatch 
is not the main source of error in A-type stars. For this error to be comparable 
to the dispersion of values obtained from different regions in the same spectrum, 
the error in the template temperature has to be $\gtrsim 2\,000$ K.

The selection of the spectral regions to be correlated depends on the quality 
and the wavelength range of the observed spectra.
Since the central wavelength of each region is used to calculate $k_1$, 
we consider 500 {\AA} as the maximum region width to minimize the influence of 
the limb darkening coefficient on $k_1$.
Typically, in an interval of  500 {\AA}, the limb darkening coefficient varies
within $\pm 0.015$, which affects the measured rotational velocity in about 
$\pm 0.25$\%. In practice, the resulting error would be significantly lower 
since it depends on the difference between coefficient for central wavelength 
(used for the calculations) and the effective coefficient, which depends on the 
spectral line distribution in the region under consideration. 
 
In addition, since the method is based on the rotational velocity as the main 
broadening factor in all spectral lines, any line strongly deviated from a 
rotational profile must be avoided. Otherwise, it will introduce an unwanted 
distortion into the CCF. Even though some of them might be known 
\textit{a priori}, e.g., \ion{He}{i} in early B stars or \ion{Ca}{ii} in late A 
to F stars, it is convenient to evaluate whether other regions introduce
a significant distortion into the CCF. 
The cross-correlation between two template spectra in different spectral regions 
can be used to detect regions that introduce significant distortions in the CCF.

\begin{figure}[t]
	\centering
\includegraphics[width=1\linewidth]{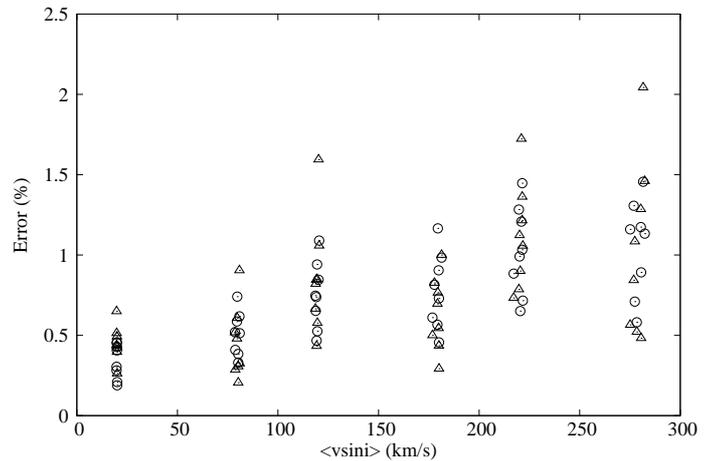} \caption{ Percentage 
error on $\langle v\sin i \rangle$ from single regions of artificially broadened 
spectra. For each velocity two values of S/N are plotted.
S/N = 100 and 200 for  $\langle v\sin i \rangle$ = 80, 180, and 280 \kms.
S/N = 100 and 70 for  $\langle v\sin i \rangle$ = 20, 120, and 220 \kms.
Triangles: Standard deviation of 10 consecutive measures (true error). 
Circles: Designated error (calibration).}
\label{fig:2}
	\end{figure}

\subsection{Single measurement error calculation}\label{sec.err}

The error assigned to a single measurement is calculated from the intensity and 
the FWHM of the CCF central maximum, and the noise present in the CCF. To calibrate 
the error formulae we used synthetic spectra artificially broadened with six different 
rotational profiles and random noise simulating S/N in the range S/N = 70 -- 200.
Then, ten spectra were generated for each pair of parameters and four regions were 
measured in each spectrum. For each region the standard deviation of the ten measures 
was adopted as the true error.

Finally, we calibrated the error as a function of the intensity I (height of the CCF 
peak), the FWHM, and the noise of the CCF (rms).
As a result, we obtained the following expression to calculate the error of a single 
region measurement: 
$\mathrm{Error (km/s)} = 4.42\ \mathrm{FWHM}^{0.520}\cdot \mathrm{rms}\cdot \mathrm{I}^{-1.08}$,
valid for spectra with S/N = 70--200, the range used for the calibration.
Figure \ref{fig:2} shows no systematic difference among real error and its calibration.

\section{Application}
\subsection{Application to A-type stars}
As part of our current research on rotational velocity, 
we have been engaged for several years in a program to measure $v\sin i$ for 
all southern A-type stars of the Bright Star Catalogue (BSC). 
This sample includes more than 800 stars . 

To test our method, we measured $\langle v\sin i \rangle$ for 251 
stars in the spectral range A1-A5\footnote{The spectral types were taken from
the SIMBAD astronomical database: 
http://simbad.u-strasbg.fr/simbad}, the results of which are presented in Table \ref{results}.
Among them 155 stars were also measured by \citet{roy02a}.
In a forthcoming paper we will publish the complete results of $v\sin i$ measurements 
for the whole sample of southern A-type stars of the BSC.

The southern A-type stars of the BSC were observed spectroscopically using the bench 
echelle spectrograph (EBASIM) fed with fibers coming from the Cassegrain focus of 
the 2.1m telescope at CASLEO. The spectrograph has been described by \citet{pintado03}.
The resolving power of the spectra varies from $\sim 30\,000$ in the blue end
to about $25\,000$ in the red end. 

The observing material was reduced using IRAF packages. 
Templates of spectral type A1--A5 with resolving power 500,000 were selected from the 
Kurucz online database \citep{ku91} to measure $v\sin i$ in objects with luminosity 
class V-IV, independently of the presence of spectroscopic peculiarities. 
For Am stars we used the template corresponding to the spectral type of the 
metallic spectrum. 

We divided our spectral range into five regions of different sizes according to the density
of spectral lines. Near the blue end of the spectra, regions of 250 {\AA} were sufficient, 
while in the red end 400 {\AA} were necessary to obtain a well-defined central maximum 
in the CCF. We avoided a rather small region between 4690 and 4730 {\AA} in all spectral 
types, because it introduces a triangular base in the CCF maximum, which makes it impossible 
to unambiguously identify the first zero in Fourier space. 
For the same reason all lines from the Balmer series were also excluded.

To simplify the calculation of the appropriate limb darkening 
coefficient for each star and wavelength region, we implemented a calibration 
$k_1(\lambda,T_\mathrm{eff})$, where $\lambda$ is the central wavelength of the spectral 
region. We used the limb darkening coefficients tabulated by \citet{claret00} and \citet{dc95}
for various photometric bands.
Then, we fitted $\varepsilon(T_\mathrm{eff})$ for our temperature range of interest,
i.e. $T_\mathrm{eff}$ = 7500\,K--11\,500\,K (spectral types B9--A9),
for $\log g$ = 4.0 and for the filters B, V, \textit{v}, \textit{b}, and \textit{y}, 
whose central wavelength are within the  range of our spectra (4000 {\AA} -- 6000 {\AA}).
The residuals of the fit for each filter lie within $\pm0.012$ (rms=0.006), which represents 
an error in $v\sin i\sim 0.15$ \%.
The final wavelength calibration was made as a function of the central wavelength of each 
filter, giving $\varepsilon(T_\mathrm{eff},\lambda)$ as a result.
By means of this calibration it is possible to calculate the $k_1$ value appropriate for 
the star's  $T_\mathrm{eff}$ and the central wavelength of spectral region.

\begin{figure}[t]
	\centering
\includegraphics[width=1\linewidth]{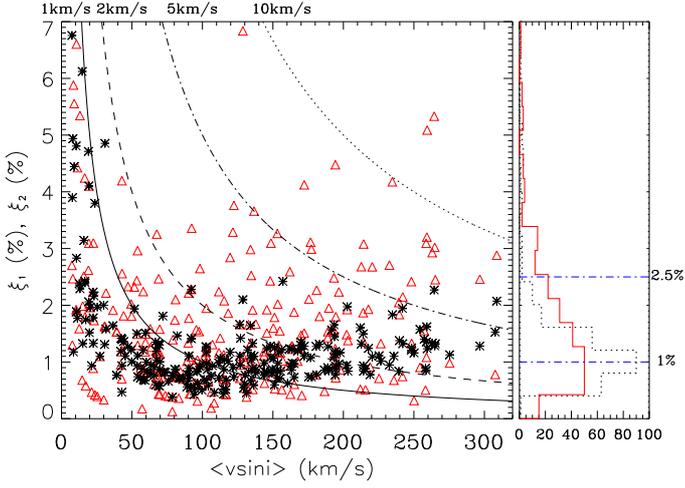} \caption{
Left: Percentage error of the 251 measured stars. Asterisks: $\xi_1$ (\%), individual 
region error indicator. Triangles: $\xi_2$ (\%), measurements dispersion indicator.
Lines represent constant values of $\xi$ in \kms. Solid: 1 \kms. Dashed: 2 \kms. 
Dot-dashed: 5 \kms. Dotted: 10 \kms.
Right: Histogram of the percentage error. Solid line: $\xi_1$ (\%). Dotted line: $\xi_2$ (\%)}
\label{fig:3}
	\end{figure}

The number of regions used to determine $v\sin i$
depends on the quality of the spectra and the wavelength range, and is 
\textit{totally independent} of $v\sin i$.
The observational material analyzed here consist of 32 spectra in the range
4000--6000 {\AA} in which we measured five regions, 116 spectra in the range 
4000--5500 {\AA} in which four regions were measured, 99 spectra in the range
3850--5000 {\AA} or in the range 4000--5250 {\AA} in which three regions were
measured, and finally four objects in which only two regions were measured owing
to excessive noise in part of the spectra.

Once $v\sin i$ and its error were  obtained  in all the regions of the same 
spectrum, the weighted mean value of $v\sin i$ was calculated, along with two 
different estimates of its uncertainty, $\xi_1$ and $\xi_2$,
following the formulae proposed by \citet[][eq.\,1]{gl00} for radial velocities.

The error $\xi_1$ is the standard deviation of the weighted mean of a
sample of $n$ uncorrelated observations with standard deviations $e_i$: 
$\xi_1 = \left[ \Sigma e_i^{-2} \right]^{-0.5}$.
On the other hand, $\xi_2$ is a generalization of the standard 
error of the mean $\sigma/\sqrt{n}$ for
a set of measurements weighted according to their individual errors.
Therefore, $\xi_2$ is calculated from the dispersion of $v\sin i$ from different 
spectral regions, while $\xi_1$ is computed from the single measurement errors 
estimated through the calibration described in Sect. \ref{sec.err}.
In general,  $\xi_1$ would be more appropriate  
when the number of measurements is small (2--3 spectral regions),
and consequently the dispersion of measurements is a less precise estimate
of the true error.
In Figure \ref{fig:3} we have plotted the values of $\xi_1$ and $\xi_2$ computed 
for each object. 
Excluding slow rotators ($v\sin i$ below 30 \kms) whose 
line widths are close to the limit imposed by the spectral resolution 
($c/R$ = 10--12 \kms), the error $\xi_1$ has an average value of 1.1\%, always 
smaller than 2.5\%.
Regarding the dispersion of the values obtained from 
different regions of the same spectrum, only three objects 
with $\langle v\sin i \rangle > 30$\,\kms\,were measured with $\xi_2$ over 5\%,
and the average error $\xi_2$ of these objects is 1.5\%.

\begin{figure}[t]
	\centering
\includegraphics[width=1\linewidth]{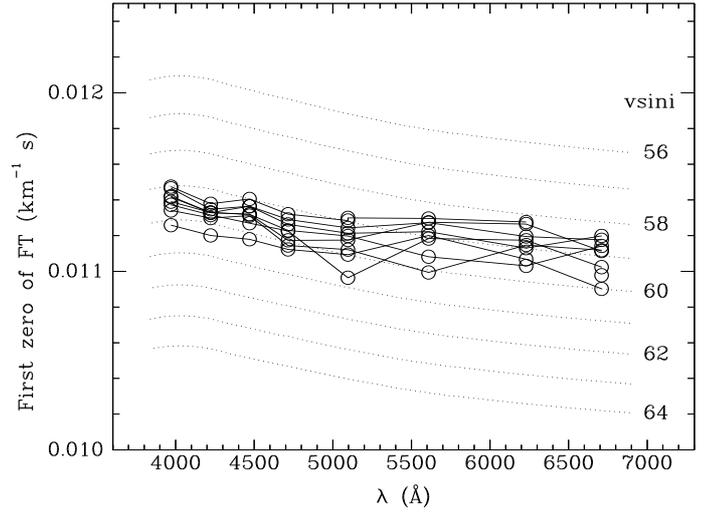}	\caption{Position of the first zero 
of the FT of the rotational profiles for 8 spectral regions of 8 spectra of the star HD77370. 
Dotted lines are theoretical lines corresponding  to different values of $v\sin i$ and have 
been calculated from the limb darkening coefficients in \citet{dc95}.
}\label{fig:raiz-lam}
	\end{figure}		
	
\subsection{Application to late type stars}
One field of application for the proposed method is the measurement of rotation in stars of
spectral type F or later, where it is not possible to measure individual lines even for low rotators.
As a test, we measured the rotational velocity of HD\,77370, an  F3\,V star
with projected rotational velocity of about 60\,\kms, high enough for all spectral lines
to appear blended.
Eight high-resolution  HARPS spectra of this star were downloaded from the ESO archive
database.
These spectra cover a spectral range 3800--6900 \AA\, with a S/N $\approx\,200$.
Rotational velocities were measured for eight spectral regions of about 200--400 \AA, using
as template a synthetic spectrum with T$_{\rm eff}$=6750\,K and solar chemical abundances.
Figure\,\ref{fig:raiz-lam} shows the measured $v\sin i$ values for the eight spectra.

\begin{figure}[t]
	\centering
\includegraphics[width=1\linewidth]{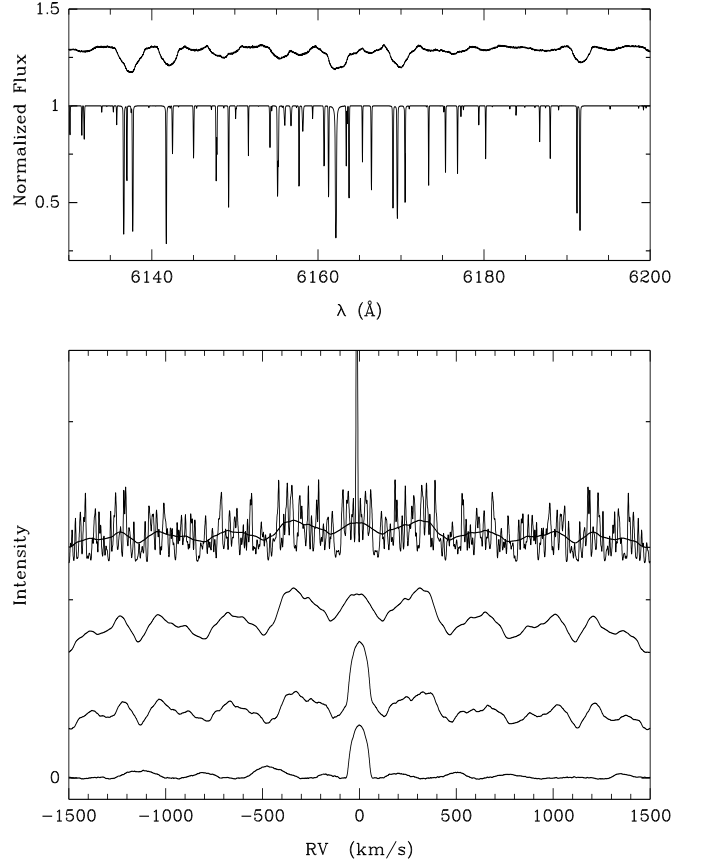} 
\caption{Example of the rotational profile calculated from a high-S/N
spectrum of only 70 \AA\,for an F-type star. 
Upper panel: Object spectrum and template spectrum. 
Lower panel: the same as Fig. 1.} \label{fig:ccf_F3}
	\end{figure}

The dependence of the position of the first zero of the FT of the rotational profile with 
wavelength -- due to the variation of the limb darkening coefficient with wavelength -- is noticeable.
The rotational velocity of each spectrum was calculated as the average of the eight spectral regions.
The mean value of the 64 measurements is 59.16\,\kms.
The standard deviation of the measurements of the 8 spectra in one single spectral region
is on average  0.41 \kms.
If we take this value as the typical uncertainty of one of the 64 
individual measurement, the uncertainty of the mean $v\sin i$  for each region should be
$\sqrt{8}$ times lower (0.14\,\kms). 
However, if we calculate the average for each of the eight spectral regions,
then the dispersion of these mean rotational velocities is
0.37\,\kms, indicating that the main source of uncertainty in these measurements
is not random, but small systematic differences between the spectral regions.
All in all, the uncertainty of the $v\sin i$ measured from a high-quality  spectrum of
an F-type star would be about 0.6\%, demonstrating that high-precision rotational
velocities can be measured even when lines are blended.
The key point in this respect is the use of the template-template correlation function
to model the subsidiary lobes of the object-template CCF during the calculation of
the rotational profile.
As an illustration, Figure \ref{fig:ccf_F3} shows the reconstruction of the rotational 
profile for a small spectral region (70 \AA) of the spectrum of HD\,77370, where all 
spectral lines are blended.

\section{Discussion}
\begin{figure}[t]
	\centering
\includegraphics[width=1\linewidth]{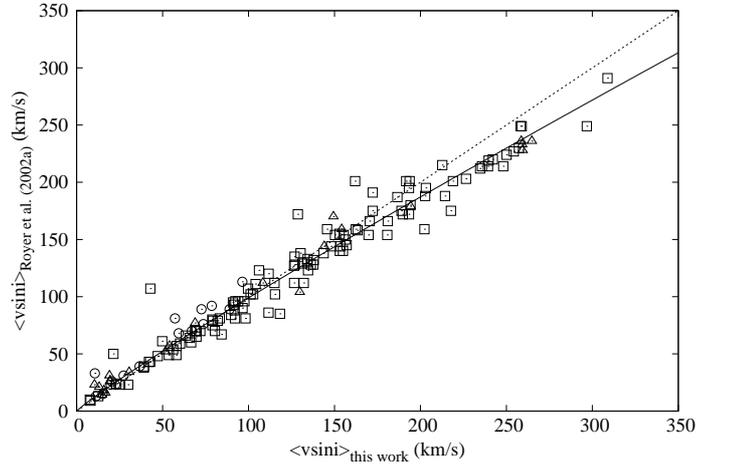} \caption{Comparison between 
our results and \citet {roy02a} for objects in common. 
Squares: \textit{normal} stars. 
Circles: \textit{peculiar} stars. 
Solid line: \textit{Normal} stars fit 
$f({\rm \langle v\sin i\rangle})=1.43\;{\rm \langle v\sin i\rangle}^{0.92}$}
\label{fig:6}
	\end{figure}

To evaluate the performance of our method in determining $v\sin i$,
we compared our results with those from \citet{roy02a}. They used FT of line profiles
to provide accurate $v\sin i$ of a large sample of A-type stars in the southern hemisphere
observed with similar resolving power ($\sim$ 28000).
Figure \ref{fig:6} shows Royer et al. values of $v\sin i$ and the ones obtained in this 
work for 155 objects in common with their sample. Even though a good agreement is found 
in stars of moderate, projected rotational velocities, a significant deviation is 
noticed for $v\sin i$ $>$ 150 \kms with an average difference of $5\pm 1$\%,
our values larger than theirs.

There are several differences between the methodology used by \citet {roy02a} and 
the present work that can account for the effect present in Figure \ref{fig:6}. 
First, they measured single lines in the range from 4200 {\AA} to 4500 {\AA}, i.e. a
region of 300 {\AA} wide which is the mean width of each region used in this work.
Second, the blending effect becomes evident through the dependence of the number of 
measured lines with $v\sin i$.
They used in average less than 3 lines on objects with $v\sin i$ $>$ 60 \kms, and over 90 \kms 
the average number of measured lines is $ \leq 2$.
Even though the dependence on the position of the first zero of the Fourier transform 
with the limb darkening coefficient was not taken into account by Royer et al., the 
adopted value $\varepsilon = 0.6$ is typical of the limb darkening coefficient in the 
range of temperatures of the objects in the sample.
However, an error of $\pm 0.01$ in $k_1$ represents an error of 1.5\% in $v\sin i$.
Thus, using a wrong $k$ value could account for 1.5\% of the deviation detected in the
 comparison.

\begin{figure}[t]
	\centering
	\includegraphics[width=1\linewidth]{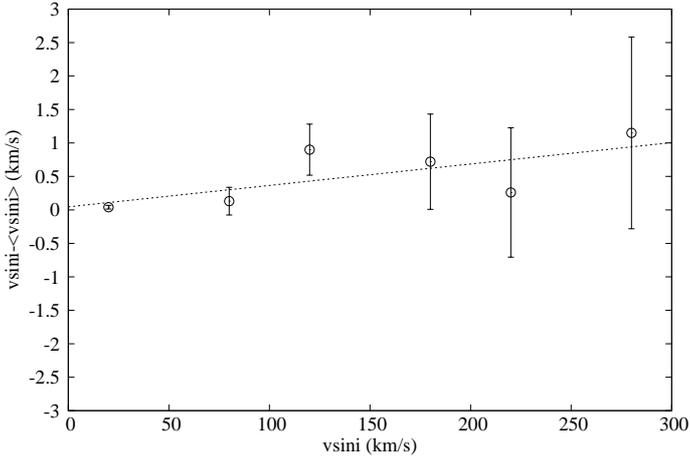}
	\caption{ Difference between real $v\sin i$ and average over ten consecutive measures 
with distinct random noise. Dotted line: 
$f({\rm \langle v\sin i\rangle})=0.0032\;{\rm \langle v\sin i\rangle}+0.0456$.}
\label{fig:7}
	\end{figure}

Finally, the main source of error is the uncertainty on the continuum position. 
\citet {roy02a} analyzed the magnitude of this effects by broadening synthetic spectra 
of $T_\mathrm{eff}$ from 7500 to 10000\,K with $v\sin i$ = 10, 50, and 100 \kms.
They found that in the coldest models with the largest $v\sin i$ the error can reach 3\%.
This uncertainty affects spectral lines giving as a result a lower value of $v\sin i$ than
the real value. 
To evaluate the intensity of this type of systematic error with our method, 
we measured $v\sin i$ in a synthetic A5V spectrum broadened by 
$v\sin i$ = 20, 80, 120, 180, 220, and 280 \kms.
Random noise was added to simulate S/N = 100, generating ten spectra with different 
noise pattern for each value of $v\sin i$. Then, four regions were used to measure $v\sin i$,
and the average over the ten measures for each region was calculated. As expected, we found 
a tendency to measure a lower value of $v\sin i$ than the actual value. 
This effect is shown in Figure \ref{fig:7}. Nevertheless, with the method proposed in 
this work the difference between measures and real values stays under 1\% 
($\langle \frac{v\sin i_{measured}}{v\sin i_{real}}\rangle_{150-300 \; \mkms} = 0.997 \pm 0.004$)
even for $v\sin i$ = 280 \kms.
In conclusion, a considerable fraction of the difference between our results and those
from \citet {roy02a} can be attributed to the effect of line blending on the continuum position.
The influence of the line blending in high rotational velocity stars is significantly less with the 
proposed method than with Fourier transform of single-line profiles,
by virtue of subtracting the secondary maximums from the CCF before calculating
the Fourier transform.
Similar results would be obtained by removing the blended lines with a similar procedure in the spectrum:
modeling the smaller lines by convolving a rotational profile with a synthetic spectrum in which 
the line of interest has been subtracted.

The main originality of the proposed method is the way in which the broadening 
function is built from the CCF, which, on the one hand, allows measurement of stars 
with blended spectral lines and, on the other, improves the S/N. 
Once the broadening function has been obtained, we determine the stellar rotation 
through the first zero of the FT. 
However, other alternative analyses are possible, such as direct fitting of the 
broadening function or Fourier-Bessel transformation \citep{pgp96}. 
 
As an illustration, we used our method in combination with the Fourier-Bessel transformation
to measure one of the objects in Piters et al. (1996) for which a HARPS spectrum was 
publicly available.
These authors applied the Fourier-Bessel transformation method to individual spectral 
lines for measuring rotation in F-type stars. Owing to line blending, stars with 
$v\sin i > 100$ \kms\, could not be measured or had errors of about 10 -- 20 \%.
We selected one of their fast rotating objects: HD\,12311, for which they 
measured 110$\pm$22 \kms.
We defined 5 spectral region between 3980 \AA\,and 6530 \AA\, obtaining
an average of 157.4 \kms with $\sigma=4.6$ \kms using our method with
the FT first zero, and 155.9 \kms\,with $\sigma=1.8$ \kms
from our broadening function but using the Fourier-Bessel transform.
Therefore, the application of  Fourier-Bessel transform method to the broadening 
function derived from the CCF would be an excellent alternative for
late-type fast-rotating stars.
In the case of slowly rotating stars, however, the Fourier-Bessel transform
is more sensitive to broadening effects that are different from rotation, so the zero of the
FT would give better results (see sect. 3.1).

\section{Summary}

We have developed a method for measuring $v\sin i$ based on the FT of the central maximum 
of the CCF. This combination provides a simple and precise solution to the line blending problem 
with medium-resolution spectra and/or spectral types later than mid-A. 
As a result, the number of useful spectral regions per object with our method is independent
of spectral type and $v\sin i$. 

The high precision of the proposed method is supported by two key features of the procedure.
The first one is to use an empirical calibration to take limb darkening effects into account in 
the position of the first zero of the FT, which is directly related to $v\sin i$. 
The second key feature is the subtraction of the subsidiary lobes of the CCF during the calculation
of the rotational broadening function.
This assures a good definition of the rotational profile even in presence of line blending. 

The $v\sin i$ value is derived from the first zero of the FT of broadening function.
However, the S/N of the resulting rotational profile is high enough to also do other 
detailed studies of the shape of the reconstructed rotational profile.

For slowly rotating stars, a lower limit  for measurable $v\sin i$ is imposed by the instrumental 
profile and the intrinsic line width, which is included twice in our broadening function owing 
to the use of a template with lines of finite width. For metallic lines this limit is 5--8 \kms.
In early-type stars the measurement is more difficult since metallic lines  become weaker
and He lines are usually not useful.
Morphological differences between template and object are also less significant in 
late-type stars. For late-A type stars or later, differences in several subtypes have
no impact on the rotation measurements.

We consider that, in accurate $v\sin i$ determinations, one significant contribution to
the errors is the adopted value for the limb darkening. Even though in this work
we use a wavelength-dependent coefficient, the continuum limb darkening might
differ significantly from the line limb darkening. Using many spectral lines in
our calculations, this difference might be averaged out, but systematic differences
between lines and continuum might result in small but systematic $v\sin i$ errors.
This is an issue that deserves further research.

We applied the proposed method to a sample of 251 A-type stars.
Measurement errors are always under 2.5\%, being 1.1\% the average error for stars with 
$\langle v\sin i\rangle$ over 30 \kms. As regards the dispersion of values measured on 
different regions across a spectral range of $\sim 1150$\,\AA\,to $2000$\,\AA\,wide, we found 
standard errors of the mean below 5\% with an average of 1.5\%.

\begin{acknowledgements}
{We thank the night assistants of CASLEO who helped during the observing procedures. 
Part of this research was supported by a grant from CONICET PIP 1113.
We gratefully acknowledge the use of ESO archival data.}
\end{acknowledgements}
\bibliographystyle{aa}

{\footnotesize
\begin{longtable}{crccc}
\caption{\normalsize Results for 251 A1--A5 stars from the Bright Star Catalogue in the 
southern hemisphere.} \\
\hline
\multicolumn{1}{c}{HD}&\multicolumn{1}{r}{$\langle vsini \rangle$}&\multicolumn{1}{c}{$\varepsilon_{1}$}&\multicolumn{1}{c}{$\varepsilon_{2}$}& \multicolumn{1}{c}{Spectral}\\
\multicolumn{1}{l}{}&\multicolumn{1}{r}{$km\; s^{-1}$}&\multicolumn{1}{c}{$km\; s^{-1}$}&\multicolumn{1}{c}{$km\; s^{-1}$}&\multicolumn{1}{c}{Type}\\
\hline 
\endfirsthead
\caption{Continued.}\\
\hline
\multicolumn{1}{c}{HD}&\multicolumn{1}{r}{$\langle vsini \rangle$}&\multicolumn{1}{c}{$\varepsilon_{1}$}&\multicolumn{1}{c}{$\varepsilon_{2}$}& \multicolumn{1}{c}{Spectral}\\
\multicolumn{1}{l}{}&\multicolumn{1}{r}{$km\; s^{-1}$}&\multicolumn{1}{c}{$km\; s^{-1}$}&\multicolumn{1}{c}{$km\; s^{-1}$}&\multicolumn{1}{c}{Type}\\
\hline
\endhead
\hline
\endfoot

319 & 60.2 & 0.6 & 0.9 & A1V \\
3719 & 53.4 & 0.8 & 1.3 & A1m \\
4772 & 163.4 & 1.4 & 2.1 & A3IV \\
6178 & 82.1 & 1.2 & 0.6 & A2V \\
6767 & 151.3 & 1.9 & 0.8 & A3IV \\
7804 & 115.1 & 1.0 & 1.3 & A3V \\
9414 & 150.3 & 1.6 & 2.4 & A2V \\
11753 & 14.7 & 0.9 & 0.01 & A3V \\
14417 & 66.6 & 0.9 & 0.6 & A3V \\
14943 & 115.0 & 0.5 & 0.8 & A5V \\
15427 & 116.8 & 0.8 & 0.5 & A2V \\
16754 & 167.6 & 2.0 & 1.7 & A1Vb \\
17168 & 49.4 & 0.6 & 1.0 & A1V \\
17254 & 126.5 & 2.1 & 0.9 & A2V \\
17566 & 90.2 & 0.8 & 2.0 & A2IV-V \\
17848 & 143.7 & 1.4 & 1.7 & A2V \\
18331 & 264.8 & 3.4 & 8.0 & A1Vn \\
18454 & 99.7 & 0.5 & 0.6 & A5IV/V \\
18543 & 53.7 & 0.5 & 0.8 & A2IV \\
18557 & 22.5 & 0.4 & 0.2 & A2m \\
18623 & 102.5 & 1.3 & 1.2 & A1V \\
20293 & 241.9 & 4.7 & 1.8 & A5V \\
20888 & 79.3 & 0.6 & 0.6 & A3V \\
21635 & 128.7 & 2.7 & 8.8 & A1V \\
21882 & 265.2 & 3.5 & 6.5 & A5V \\
21997 & 69.8 & 0.7 & 0.8 & A3IV/V \\
22243 & 189.0 & 2.4 & 2.9 & A2V \\
22634 & 68.9 & 0.9 & 1.7 & A3V \\
23055 & 92.1 & 2.1 & 2.1 & A3IV/V \\
23281 & 78.6 & 0.3 & 0.03 & A5m \\
23738 & 193.3 & 3.2 & 4.8 & A3V \\
24554 & 9.0 & 0.4 & 0.5 & A2V \\
25371 & 135.7 & 1.2 & 0.7 & A2V \\
27411 & 20.4 & 0.4 & 0.4 & A3m... \\
27490 & 157.1 & 3.8 & 1.6 & A3V \\
27861 & 194.3 & 2.3 & 8.7 & A2V \\
28312 & 162.4 & 1.5 & 0.7 & A3V \\
28763 & 102.7 & 1.2 & 1.3 & A2/A3V \\
30127 & 193.3 & 2.1 & 3.0 & A1V \\
30321 & 134.1 & 2.2 & 4.0 & A2V \\
30422 & 127.0 & 1.8 & 1.2 & A3IV \\
31093 & 196.8 & 1.9 & 4.4 & A1Vn \\
31739 & 143.7 & 1.1 & 1.6 & A2V \\
32667 & 144.0 & 1.2 & 0.9 & A2IV \\
37306 & 148.1 & 1.9 & 3.6 & A2V \\
38678 & 258.7 & 2.2 & 8.0 & A2IV-Vn: \\
39421 & 254.3 & 4.0 & 2.3 & A2Vn \\
39789 & 202.7 & 4.0 & 3.9 & A3IV \\
41759 & 212.7 & 2.8 & 5.7 & A1V \\
41841 & 58.2 & 0.4 & 0.5 & A2V \\
42824 & 134.4 & 1.5 & 1.3 & A2V \\
43319 & 73.7 & 0.6 & 0.3 & A5IVs \\
43847 & 11.3 & 0.2 & 0.5 & A2Vm... \\
43940 & 258.2 & 2.3 & 1.3 & A2V \\
48915 & 16.7 & 0.4 & 0.3 & A1V \\
50445 & 94.9 & 0.5 & 0.6 & A3V \\
50747 & 84.3 & 0.6 & 0.7 & A4IV \\
51055 & 42.8 & 0.2 & 0.3 & A2V \\
53811 & 63.0 & 0.5 & 1.0 & A4IV \\
55185 & 175.5 & 1.3 & 3.7 & A2V \\
55595 & 154.9 & 1.1 & 2.4 & A5IV-V \\
56405 & 145.7 & 1.8 & 2.8 & A1V \\
57240 & 47.2 & 0.6 & 1.1 & A1V \\
62864 & 71.9 & 1.3 & 0.5 & A2V \\
65456 & 39.2 & 0.3 & 0.6 & A2Vv \\
65810 & 239.3 & 2.0 & 4.9 & A1V \\
66210 & 30.9 & 1.5 & 0.5 & A2V \\
68862 & 91.9 & 0.5 & 1.0 & A3V \\
69665 & 98.2 & 0.9 & 2.6 & A1V \\
70340 & 10.6 & 0.3 & 0.2 & A2Vpn:EuSrCr:Si\\
70612 & 104.1 & 0.6 & 0.7 & A3V \\
71267 & 19.8 & 0.3 & 0.4 & A3m \\
71688 & 132.1 & 1.6 & 0.7 & A1/A2V \\
71815 & 39.1 & 0.3 & 0.6 & A1/A2V \\
72660 & 8.1 & 0.4 & 0.2 & A1V \\
72968 & 15.9 & 0.5 & 0.4 & A1pSrCrEu\\
73997 & 193.7 & 2.8 & 2.2 & A1Vn \\
74190 & 59.4 & 0.4 & 0.7 & A5m \\
74341 & 97.6 & 0.9 & 1.5 & A3V \\
74879 & 63.5 & 0.6 & 0.8 & A3IV/V \\
75171 & 96.5 & 0.6 & 0.6 & A4V \\
75630 & 163.7 & 1.6 & 2.1 & A2/A3IV \\
75737 & 16.5 & 0.4 & 0.7 & A4m \\
75926 & 194.5 & 2.2 & 1.4 & A1Vn \\
76483 & 68.9 & 0.4 & 0.5 & A3IV \\
78045 & 30.6 & 0.4 & 0.5 & kA3hA5mA5v \\
78676 & 56.1 & 0.4 & 0.5 & A4IV \\
78922 & 108.3 & 0.5 & 0.3 & A4IV-V \\
79193 & 10.4 & 0.5 & 0.2 & A3III:m+A0V:\\
80007 & 145.7 & 1.2 & 2.2 & A2IV \\
80447 & 15.1 & 0.2 & 0.3 & A2Vs \\
80951 & 23.7 & 0.9 & 0.1 & A1V \\
81157 & 36.6 & 0.6 & 0.7 & A3IVs... \\
81309 & 10.6 & 0.2 & 0.7 & A2m \\
82068 & 258.9 & 3.5 & 2.8 & A3Vn \\
82446 & 54.0 & 0.6 & 1.6 & A3V \\
83520 & 149.6 & 1.3 & 1.8 & A2,5V \\
83523 & 134.4 & 0.9 & 1.8 & A2V \\
85558 & 134.6 & 1.4 & 3.4 & A2V \\
86266 & 186.6 & 1.6 & 0.9 & A4V \\
86301 & 218.9 & 1.7 & 2.0 & A4V \\
88024 & 92.3 & 1.3 & 3.1 & A2V \\
88372 & 259.2 & 4.2 & 8.3 & A2Vn \\
88522 & 23.1 & 0.5 & 0.4 & A1V \\
88842 & 78.9 & 0.6 & 0.3 & A3IV-V \\
88955 & 103.6 & 0.9 & 1.1 & A2Va \\
89263 & 67.2 & 0.8 & 0.5 & A5V \\
89816 & 214.3 & 2.0 & 6.6 & A4IV/V \\
90630 & 90.9 & 0.7 & 0.4 & A2,5V \\
90874 & 65.7 & 0.5 & 0.4 & A2V \\
91790 & 111.7 & 0.8 & 1.3 & A5IV/V \\
93397 & 94.0 & 0.4 & 0.9 & A3V \\
93742 & 49.8 & 0.5 & 0.9 & A2IV \\
93903 & 19.4 & 0.4 & 0.6 & A3m \\
93905 & 118.3 & 1.2 & 2.2 & A1V \\
94985 & 181.0 & 1.4 & 4.1 & A4V \\
95370 & 111.6 & 1.1 & 2.5 & A3IV \\
96124 & 156.3 & 1.9 & 0.8 & A1V \\
96146 & 7.7 & 0.3 & 0.03 & A2V \\
96441 & 130.3 & 1.0 & 1.3 & A1V \\
96723 & 25.4 & 0.5 & 0.1 & A1V \\
103101 & 42.1 & 0.5 & 0.4 & A2V \\
103266 & 165.1 & 1.6 & 1.0 & A2V \\
104039 & 12.6 & 0.3 & 0.2 & A1IV/V \\
105776 & 153.1 & 1.1 & 5.0 & A5V \\
105850 & 126.8 & 1.2 & 1.4 & A1V \\
106819 & 79.0 & 0.6 & 1.1 & A2V \\
107070 & 202.3 & 1.9 & 3.0 & A5Vn \\
108107 & 242.1 & 3.2 & 5.2 & A1V \\
108925 & 188.9 & 1.7 & 2.7 & A3V \\
109074 & 89.9 & 0.6 & 1.2 & A3V \\
109704 & 155.4 & 1.3 & 1.8 & A3V \\
109787 & 296.8 & 3.8 & 7.3 & A2V \\
111588 & 60.6 & 0.5 & 0.5 & A5V \\
114330 & 7.4 & 0.5 & 0.2 & A1IVs+... \\
114576 & 162.0 & 2.0 & 4.2 & A5V \\
115892 & 90.3 & 1.0 & 1.6 & kA15hA3mA3va \\
116061 & 194.2 & 1.5 & 2.4 & A2/A3V \\
116197 & 239.7 & 3.3 & 4.4 & A4V \\
117150 & 220.4 & 2.4 & 5.9 & A1V \\
117558 & 169.8 & 2.2 & 2.1 & A1V \\
118098 & 232.6 & 2.2 & 6.3 & A3V \\
119938 & 69.7 & 0.4 & 0.3 & A3m... \\
122958 & 172.2 & 1.9 & 7.1 & A1/A2V \\
123998 & 17.2 & 0.3 & 0.1 & A2m \\
124576 & 126.5 & 1.4 & 2.1 & A1V \\
125283 & 257.3 & 3.6 & 4.0 & A2Vn \\
126367 & 65.8 & 0.6 & 0.7 & A1/A2V \\
126722 & 101.2 & 0.7 & 1.1 & A2IV \\
127716 & 96.4 & 0.8 & 0.3 & A2IV \\
130841 & 59.6 & 0.9 & 0.8 & kA2hA5mA4Iv-v \\
132219 & 42.9 & 0.4 & 1.8 & A4V+,,, \\
133112 & 96.4 & 0.6 & 1.0 & A5m \\
134482 & 160.4 & 1.0 & 1.5 & A3IV \\
134967 & 259.4 & 3.2 & 13.2 & A2V \\
135235 & 75.9 & 0.5 & 0.3 & A3m \\
135379 & 68.5 & 0.9 & 0.8 & A3Va \\
137015 & 119.6 & 1.4 & 3.4 & A2V \\
137333 & 44.1 & 0.6 & 0.8 & A2V \\
138413 & 21.4 & 0.2 & 0.1 & A2IV \\
138965 & 102.7 & 1.0 & 2.0 & A1V \\
141413 & 116.6 & 0.6 & 0.8 & A5IV \\
142139 & 89.2 & 0.7 & 0.7 & A3V \\
142445 & 137.7 & 1.1 & 1.4 & A3V \\
142629 & 43.0 & 0.6 & 1.1 & A3V \\
143101 & 86.1 & 0.6 & 0.6 & A5V \\
145570 & 42.3 & 0.6 & 0.5 & A3IV \\
145607 & 217.7 & 2.7 & 4.2 & A4V \\
145689 & 106.4 & 0.8 & 0.5 & A4V \\
146667 & 225.2 & 2.5 & 1.6 & A3Vn \\
148367 & 19.5 & 0.8 & 0.8 & A3m \\
150573 & 235.8 & 2.6 & 1.6 & A4V \\
150894 & 136.2 & 0.8 & 0.9 & A3IV \\
151676 & 149.4 & 1.2 & 1.1 & A3V \\
152127 & 31.6 & 0.5 & 0.7 & A2Vs \\
153053 & 102.8 & 0.7 & 1.0 & A5IV-V \\
154310 & 263.0 & 3.5 & 7.7 & A2IV \\
154418 & 72.6 & 0.6 & 0.6 & A1m,,, \\
154494 & 114.8 & 0.6 & 0.8 & A4IV \\
154895 & 121.6 & 1.1 & 4.0 & A1V+F3V\\
155259 & 214.0 & 3.2 & 4.0 & A1V \\
159492 & 54.1 & 0.4 & 0.7 & A5IV-V \\
160613 & 112.6 & 1.2 & 1.6 & A2Va \\
164577 & 194.6 & 1.8 & 3.5 & A2Vn \\
165189 & 129.8 & 0.8 & 0.7 & A5V \\
165190 & 129.2 & 0.7 & 1.4 & A5V \\
166393 & 191.7 & 3.0 & 3.0 & A2V \\
166960 & 27.3 & 0.3 & 0.3 & A2m \\
167564 & 153.0 & 1.3 & 1.8 & A4V \\
169853 & 22.6 & 0.3 & 0.7 & A2m \\
170384 & 131.3 & 0.6 & 0.5 & A3V \\
170479 & 67.6 & 0.7 & 2.2 & A5V \\
170642 & 177.3 & 1.3 & 5.3 & A3Vn \\
171856 & 106.1 & 0.8 & 0.2 & A5IV \\
172555 & 116.4 & 0.6 & 0.5 & A7V \\
172777 & 148.6 & 1.6 & 4.6 & A2Vn \\
173715 & 122.3 & 0.9 & 4.6 & A3V \\
175638 & 151.1 & 0.9 & 2.7 & A5V \\
175639 & 214.3 & 1.8 & 2.9 & A5Vn \\
176687 & 68.9 & 0.4 & 0.3 & A2,5Va \\
176984 & 30.0 & 0.6 & 0.1 & A1V \\
178253 & 203.2 & 1.7 & 2.8 & A2Va \\
180482 & 80.4 & 0.7 & 1.4 & A3IV \\
181383 & 180.7 & 2.3 & 1.4 & A2V \\
183545 & 226.5 & 2.3 & 3.0 & A2V \\
184552 & 13.1 & 0.3 & 0.7 & A1m \\
184586 & 197.7 & 2.3 & 1.7 & A1V \\
187421 & 170.0 & 2.3 & 2.5 & A2V \\
187653 & 248.2 & 3.0 & 6.0 & A4V \\
188899 & 69.7 & 0.6 & 1.1 & A2V \\
189118 & 45.2 & 0.5 & 0.6 & A4/A5IV \\
189388 & 248.4 & 3.1 & 3.3 & A2,5V \\
189741 & 137.9 & 1.2 & 2.3 & A1IV \\
197725 & 153.4 & 1.5 & 3.1 & A1V \\
198001 & 115.4 & 1.2 & 2.4 & A1,5V \\
198160 & 183.0 & 2.4 & 3.3 & A2,5IV-V \\
198161 & 176.7 & 1.9 & 5.5 & A2,5IV-V \\
198529 & 275.4 & 3.1 & 2.7 & A2Vn \\
198743 & 51.7 & 1.0 & 1.2 & A3m \\
199475 & 212.0 & 3.4 & 1.1 & A2V \\
201906 & 264.4 & 6.0 & 14.1 & A1V \\
202730 & 135.6 & 1.8 & 2.6 & A5Vn: \\
203562 & 58.0 & 0.7 & 1.3 & A3V \\
204394 & 132.7 & 1.1 & 2.2 & A1V \\
204854 & 107.1 & 1.0 & 1.4 & A2IVv \\
205765 & 172.2 & 2.2 & 2.6 & A2V \\
205811 & 102.6 & 1.3 & 1.3 & A2V \\
207155 & 136.6 & 1.5 & 5.0 & A2V \\
208321 & 215.6 & 1.9 & 1.1 & A2Vn \\
208565 & 308.8 & 6.4 & 8.9 & A2Vnn \\
210049 & 307.7 & 4.7 & 2.4 & A1,5IVn \\
210739 & 189.7 & 1.6 & 3.7 & A3V \\
212728 & 250.1 & 2.9 & 0.8 & A4V \\
213884 & 165.0 & 1.2 & 3.1 & A5V \\
214085 & 163.4 & 2.0 & 2.0 & A3Vn \\
214150 & 80.9 & 0.7 & 1.7 & A1V \\
214484 & 8.5 & 0.1 & 0.5 & A2Vp \\
215729 & 170.5 & 2.1 & 0.7 & A2V \\
216627 & 83.2 & 0.6 & 0.3 & A3V \\
216900 & 67.0 & 0.4 & 0.8 & A3Vs \\
216956 & 91.6 & 0.5 & 0.4 & A4V \\
217498 & 91.9 & 1.3 & 2.1 & A2V \\
218108 & 194.5 & 1.7 & 1.9 & A3,5V \\
219402 & 154.2 & 1.8 & 2.0 & A3V \\
221675 & 57.2 & 0.5 & 0.1 & A2m \\
221760 & 22.4 & 0.5 & 0.5 & A2VpSrCrEu\\
222602 & 234.7 & 4.3 & 9.8 & A3Vn \\
223438 & 89.0 & 0.5 & 0.9 & A5m \\
223466 & 69.5 & 0.9 & 0.6 & A3V \\
223991 & 19.1 & 0.9 & 0.5 & A2V+F2V\\
224361 & 95.6 & 0.8 & 1.5 & A1IV \\

\end{longtable}
}

\end{document}